\documentclass[lineno]{article}

\usepackage[utf8]{inputenc}
\usepackage{graphicx}
\usepackage{amsfonts}
\usepackage{ulem}
\usepackage[square,numbers]{natbib}
\usepackage{subcaption}
\usepackage{url,lineno}
\usepackage{tikz}
\usepackage{float}
\usepackage[labelfont=bf, font=scriptsize]{caption}
\usepackage{placeins}
\usepackage{authblk}
\usepackage[a4paper, total={6.3in, 10in}]{geometry}
\usepackage[colorlinks=true, citecolor=blue, filecolor=blue, linkcolor=black]{hyperref}
\usepackage{bbm}
\usepackage{color}
\usepackage{epstopdf}
\usepackage{stmaryrd}
\usepackage{comment}
\usepackage{multicol}
\usepackage{amsmath,amssymb} 
\usepackage{float}
\usepackage{color}
\usepackage{mathtools}
\usepackage[T1]{fontenc}    
\usepackage{url}            
\usepackage{booktabs}       
\usepackage{amsfonts}       
\usepackage{nicefrac}       
\usepackage{microtype}      
\usepackage{lipsum}
\usepackage{xcolor,colortbl}
\usepackage{amsmath}
\usepackage[square,numbers]{natbib}
\usepackage{lastpage,fancyhdr}

\newcommand{\be}{\begin{equation}}
\newcommand{\ee}{\end{equation}}
\newcommand{\bea}{\begin{eqnarray}}
\newcommand{\eea}{\end{eqnarray}}

\definecolor{Gray}{gray}{0.95}
\newcolumntype{Q}[1]{>{\columncolor{Gray}}p{#1}}

\definecolor{lime}{HTML}{A6CE39}
\DeclareRobustCommand{\orcidicon}{%
	\begin{tikzpicture}
	\draw[lime, fill=lime] (0,0) 
	circle [radius=0.16] 
	node[white] {{\fontfamily{qag}\selectfont \tiny ID}};
	\draw[white, fill=white] (-0.0625,0.095) 
	circle [radius=0.007];
	\end{tikzpicture}
	\hspace{-2mm}
}

\foreach \x in {A, ..., Z}{%
	\expandafter\xdef\csname orcid\x\endcsname{\noexpand\href{https://orcid.org/\csname orcidauthor\x\endcsname}{\noexpand\orcidicon}}
}

\begin{document}  

\title{\textbf{Event-based spatiotemporal networks for modelling emergent phenomena in complex systems}}

\author[1]{Matthijs Romeijnders\orcidA{}\footnote{Email: m.c.romeijnders@uu.nl}}
\affil[1]{Department of Information and Computing Sciences, Utrecht University, The Netherlands}
\author[2]{Michiel van Boven\orcidB{}\footnote{Email: r.m.vanboven-2@umcutrecht.nl}}
\affil[2]{Julius Center for Health Sciences and Primary Care, Utrecht University, Utrecht, Netherlands}
\author[3]{Francesco Corman\orcidC{}\footnote{Email: francesco.corman@ivt.baug.ethz.ch}}
\affil[3]{Institute for Transport Planning and Systems, ETH Z\"urich,  Z\"urich, Switzerland}
\author[4,5,6]{Carl D. Modes\orcidD{}\footnote{Email: modes@mpi-cbg.de}}
\affil[4]{Max-Planck Institute for Molecular Cell Biology and Genetics, Dresden, Germany}
\affil[5]{Center for Systems Biology Dresden, Dresden, Germany}
\affil[6]{Cluster of Excellence, Physics of Life, TU Dresden, 01307 Dresden, Germany}
\author[7,8,9]{Phillip P. A. Staniczenko\orcidE{}\footnote{Email: pstaniczenko@brooklyn.cuny.edu}}
\affil[7]{Department of Biology, Brooklyn College, City University of New York, Brooklyn, New York, USA}
\affil[8]{Biology PhD Program, CUNY Graduate Center, New York, USA}
\affil[9]{Division of Invertebrate Zoology, American Museum of Natural History, New York, USA}
\author[1]{Debabrata Panja\orcidE{}\footnote{Corresponding author: d.panja@uu.nl}}


\maketitle 

\begin{abstract}
\noindent Complex systems display emergent phenomena that vary significantly across spatial and temporal scales. These variations originate from fine-grained system processes, yet arriving at macroscopic dynamics from micro-level data --- particularly when large, high-resolution datasets are available --- remains a persistent challenge. Here we develop event-based spatiotemporal networks, a computational modelling framework that encodes system processes as discrete events anchored in space and time. Event-based spatiotemporal networks offer a unified, flexible and efficient approach to generate emergent behaviour in complex systems across space and time from these events. We demonstrate the effectiveness of event-based spatiotemporal networks through two illustrative real-world applications. First, following a local outbreak of a novel respiratory pathogen in the Netherlands, spatiotemporal networks enable fine-grained tracking of transmission routes and infection patterns through space and time. Second, we use spatiotemporal networks to model propagation of delays in a public transportation system (S-bahn) around Z\"urich, Switzerland. We also discuss broader uses of event-based spatiotemporal networks in fields like developmental biology and community ecology, where focusing on events rather than static system states can improve data analysis, simulation, and collection strategies.
\end{abstract}

\vspace{5mm}
\noindent Emergent phenomena such as traffic bottlenecks \cite{ambuhl2023understanding}, spreading patterns of infectious diseases \cite{vespig2021}, and flocking in birds and other animals \cite{viscek2012} are characteristics of  complex systems. These are systems composed of many mobile actors that are heterogeneously distributed across space and time, where the dynamics of system-level (macro-scale) properties emerge from fine-grained (micro-scale) constitutive system processes --- i.e., processes that give rise to the organised existence of the emergent phenomena. The increasing availability of large datasets with highly resolved spatial, temporal and actor information on these constitutive processes raises possibilities for unravelling how micro-scale heterogeneities shape macro-scale behaviour of complex systems in space and time, but also brings about challenges in balancing model details with computational feasibility.

\vspace{2mm}
\noindent An often-used approach for managing model complexity is to compartmentalise space and time into intermediate-scale building blocks that can be considered large compared to the micro-scales of the constitutive processes but small compared to the macro-scales of emergent phenomena. This approach is justifiable when these processes occur at scales that are orders of magnitude smaller than the emergent phenomenon of interest. It assumes that fine-grained spatial, temporal and actor information within compartments can be safely averaged out, reducing the many forms of actor heterogeneities to a much smaller number of general compartmental variables, a process that is usually referred to as coarse-graining \cite{voth2013}. Further, when additional mechanistic formulations of how compartmental variables impact each other across compartments can be constructed, the coarse-graining step --- with the use of the spatiotemporal compartments as the fundamental units of space and time --- makes way for developing a dynamical system approach to describe future system trajectories at the macro-scale. The use of partial differential equations to describe these dynamical systems, thereby opening up future system trajectories to analytical or numerical tractability, has proved to be quite successful descriptions of emergent phenomena in complex systems. Examples of this approach are neural mass models in neuroscience \cite{wc1972}, (meta)population models in ecology \cite{hanski1997}, and compartmental rate (partial) differential equation models in epidemiology \cite{diekmann2013}.

\vspace{2mm}
\noindent Even when compartments cannot be easily defined because there is no clear delineation of scales, they are still often constructed to facilitate mathematical treatments and reduce computation time for simulations. Despite the benefits, doing so runs the risk of missing out on important relationships among micro-scale heterogeneities and their macro-scale implications (a practice often referred to as uncontrolled approximation \cite{mann2022}). The dynamics of traffic jams exemplify this situation since individual car movements are close to the spatial and temporal scales of resulting bottlenecks, and the dynamics of traffic jams are well-known to be highly sensitive to specific traffic configurations and the accompanying behavioural responses of drivers. As these intricate constitutive processes at the micro-scale are consequential for understanding the overall system dynamics, the conceptual and practical benefits of taking a partial differential equation approach in such cases are limited. For data-rich complex systems whose constitutive processes at the micro-scale are inherently heterogeneous across space and time scales and cannot be easily compartmentalised, a scalable computational framework that combines space, time, and actor heterogeneities is necessary.

\section*{Results}

\subsection*{Event-based spatiotemporal networks}

\noindent We introduce event-based spatiotemporal networks (EBSTNs) as a computational modelling framework that flexibly and efficiently retains all relevant spatial, temporal and actor information in the constitutive system processes at the micro-scale. We refer to the spatiotemporal ``containers'', where these constitutive processes occur, as {\it events}. The processes in the four application areas discussed in this paper are: (i) transmission of a pathogen to a susceptible individual in infectious-disease epidemiology, which requires a gathering of humans as actors; (ii) the operation of a train in public transport, which involves an assembly of the operating crew and vehicle(s); (iii) spatially localised topological transitions in a cellular matrix in developmental biology, realised through interactions among participating cells via biochemical and mechanosensory signals; and (iv) feeding, competition, pollination, and related processes in community ecology that entail interactions among species' members as actors. In this way, events provide the spatiotemporal anchor points for the micro-scale constitutive processes of a complex system (an example in the next subsection). Underpinned by the relevant interactions among the system actors, events then motivate the development of EBSTNs as the natural computational framework for modelling macro-scale phenomena in space and time that emerge from heterogeneous micro-scale  constitutive system processes.

\begin{figure}[!h]
\centering   \includegraphics[width=\linewidth]{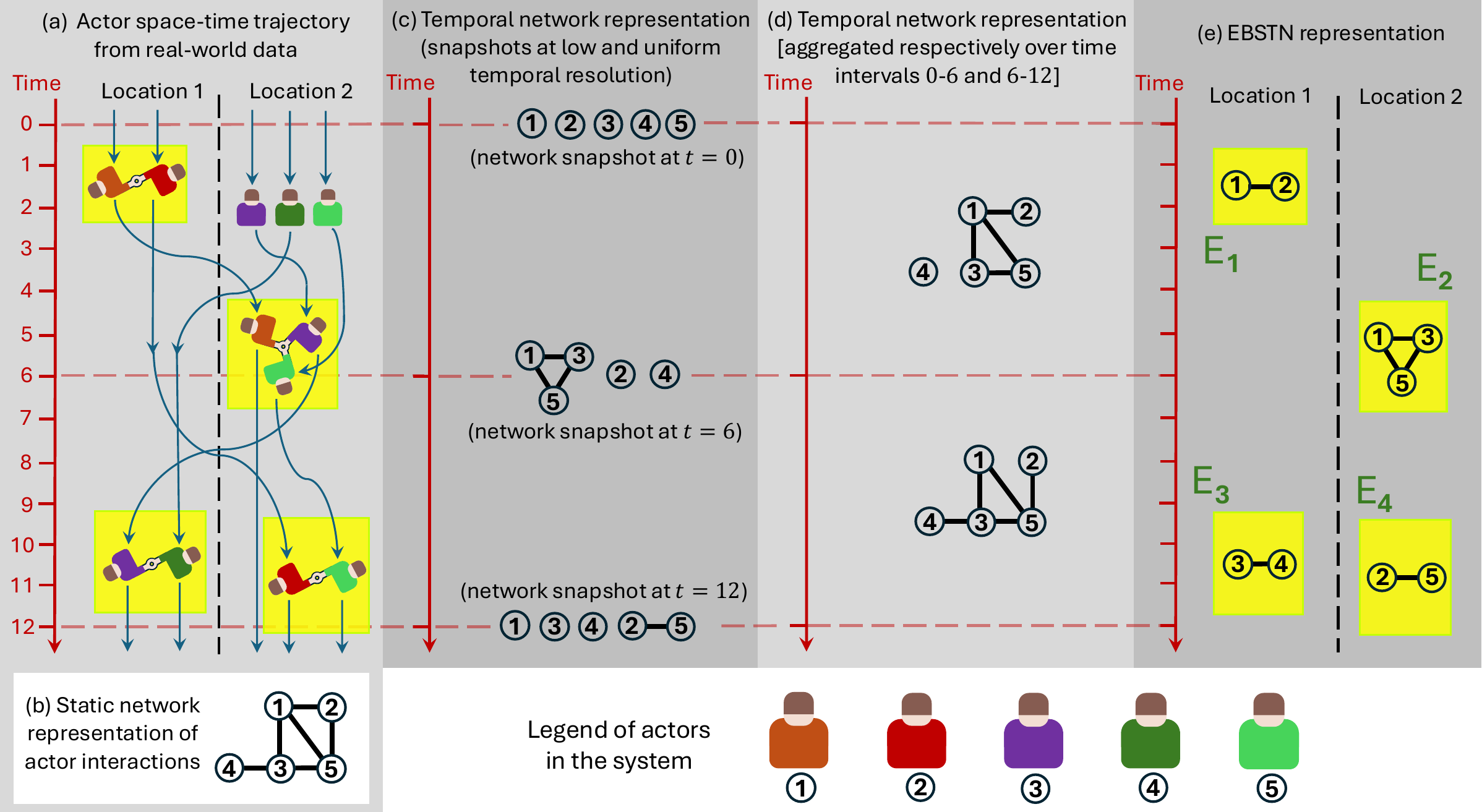}
\caption{{\bf From real-world data to network representations.} Panel (a): Actor dynamics in a schematic real-world system (see legend of actors in colour and alphanumeric codes). Yellow rectangles denote locations, timings and durations of actor interactions: the top and the bottom boundaries of every rectangle respectively denote the starting and the ending time of actor interactions. Shown in panel (b) is a static network representation of it is obtained when space and time are both abstracted away; it loses all information on the times, locations, sequence and durations of actor interactions. In panel (c) we show a version of temporal network formed in multiple layers, each layer corresponds to a snapshot. The snapshots are taken at a ``low'' resolution for using the data feasibly in models, at the cost of missing actor interactions. In panel (d) we show a version of temporal network formed in multiple layers too, but this time upon aggregating the data over the same uniform time intervals as in (c), demonstrating how aggregations can mask the true nature of the actor interactions. Finally, in panel (e) we show the corresponding event-based spatio-temporal network (EBSTN), with every interaction being denoted as an {\it event\/}, labelled $E_1$ through $E_4$, matching to the yellow rectangles in panel (a), and showcasing that in the EBSTN formulation there is no loss of relevant information on the true nature of actor interactions. Note in panel (e) that the actor movements across events are not explicitly shown. Panels (a) and (e) have been adapted from Ref. \cite{panja2025}.}
\label{fig1}
\end{figure}

\vspace{2mm}
\noindent By construction, EBSTNs take a {\it dynamic\/} network approach for modelling and analysis of emergent dynamical behaviour of complex systems. Since the establishment of network science as a research field, time-invariant (static) networks --- collections of actors connected by fixed links --- have been, and continue to be, widely used for this purpose \cite{barratbook2012}. In recent years, however, it has become increasingly clear that static networks fall short in modelling the dynamics of many complex systems, as emergent dynamics in space and time can critically depend, through dynamic interdependencies, on the timing and sequence of actor interactions \cite{dekker2022a,dekker2022b}. Temporal networks (TNs) \cite{Holme2012} --- the current best dynamic network framework --- address this by encoding when interactions occur and in which order, allowing links among actors to change over time. With actors as nodes and their interactions as links, TNs preserve the {\it actor-centric\/} nature of static networks, while enriching it through system-wide snapshots or temporally aggregated interaction layers, typically constructed at uniform time intervals \cite{socpatt} [Fig. \ref{fig1}(c–d)]. However, the actor-centric approach can quickly become overwhelmed when tasked with incorporating large number of actors and large amounts of finely-resolved actor-interaction data. EBSTNs overcome this limitation through an {\it event-centric\/} --- topologically dual --- representation, where the (spatiotemporally-anchored) events and the actor movements across them can be construed as network nodes and links respectively [Fig. 1(e)]. This duality makes the (causal) interdependencies explicit rather than implicit (see Discussion).

\vspace{2mm}
\noindent EBSTNs encode four distinct types of information on the constitutive processes at every event: where (process location), when (process timing), who (which actors are involved) and how long (process duration). The key to the advantage of the EBSTN framework, as we elucidate in the next subsection, is twofold: (1) unique event encoding requires less computer memory, and (2) it encodes actor movement implicitly in the event structure, eliminating the need to iterate over individual actor positions at each simulation time step. Indeed, shifting to event-based encoding --- where, when, who, and how long --- encapsulates all the information required for unravelling how micro-scale heterogeneities shape macro-scale behaviour of complex systems in space and time. In Fig. \ref{fig1}(e) we illustrate the construction of an EBSTN from real-world data of an imagined complex system as a collection of events and how that differs from static, or versions of temporal networks. In contrast, EBSTNs allow researchers to retain flexibility in choosing an appropriate (nonuniform) sampling interval, and potentially assess the consequences of that choice. Crucially, with actor movements connecting events, EBSTNs naturally (a) incorporate empirical constraints on actor mobility that limit which actors can participate in which event, and (b) build in constitutive process interdependencies across space and time that dictate the dynamics of emergent phenomena. 

\vspace{2mm}
\noindent We note that there are mentions of spatiotemporal networks in the existing literature \cite{george2013,Williams2016,zhang2020}.These are extensions of spatial networks \cite{barthelemy2011}. In these spatiotemporal networks nodes remain fixed in space just as in spatial networks, while the temporal dimension captures flows (e.g., information) along links. These spatiotemporal networks  focus on topological properties of spatial networks and their impact on flows. They are fundamentally unrelated to dynamic interactions among mobile actors we focus on here. 

\vspace{2mm}
\noindent We showcase two applications of EBSTNs: modelling the outbreak of a novel respiratory pathogen in the Netherlands, and the development of delays in the Z\"urich S-bahn system (Sihltal-Zürich-Uetliberg Bahn, or SZU). In infectious disease epidemiology, coarse-graining data into compartments and invoking partial differential equations is commonplace, but taking this approach would miss nuances such as specific transmission routes and heterogeneous infection patterns through space and time. The extent of spatial and temporal non-uniformities in the S-bahn scheduling data means that a partial differential equation approach cannot even be constructed, requiring a modelling approach with finer scale resolution for which EBSTNs are particularly suitable. We also discuss applications of EBSTNs in developmental biology and community ecology that could explain observed spatiotemporal variations at system-level scales.

\subsection*{From constitutive processes  to events to computational efficiency}

\begin{figure}[!h]
\centering
\includegraphics[width=0.7\linewidth]{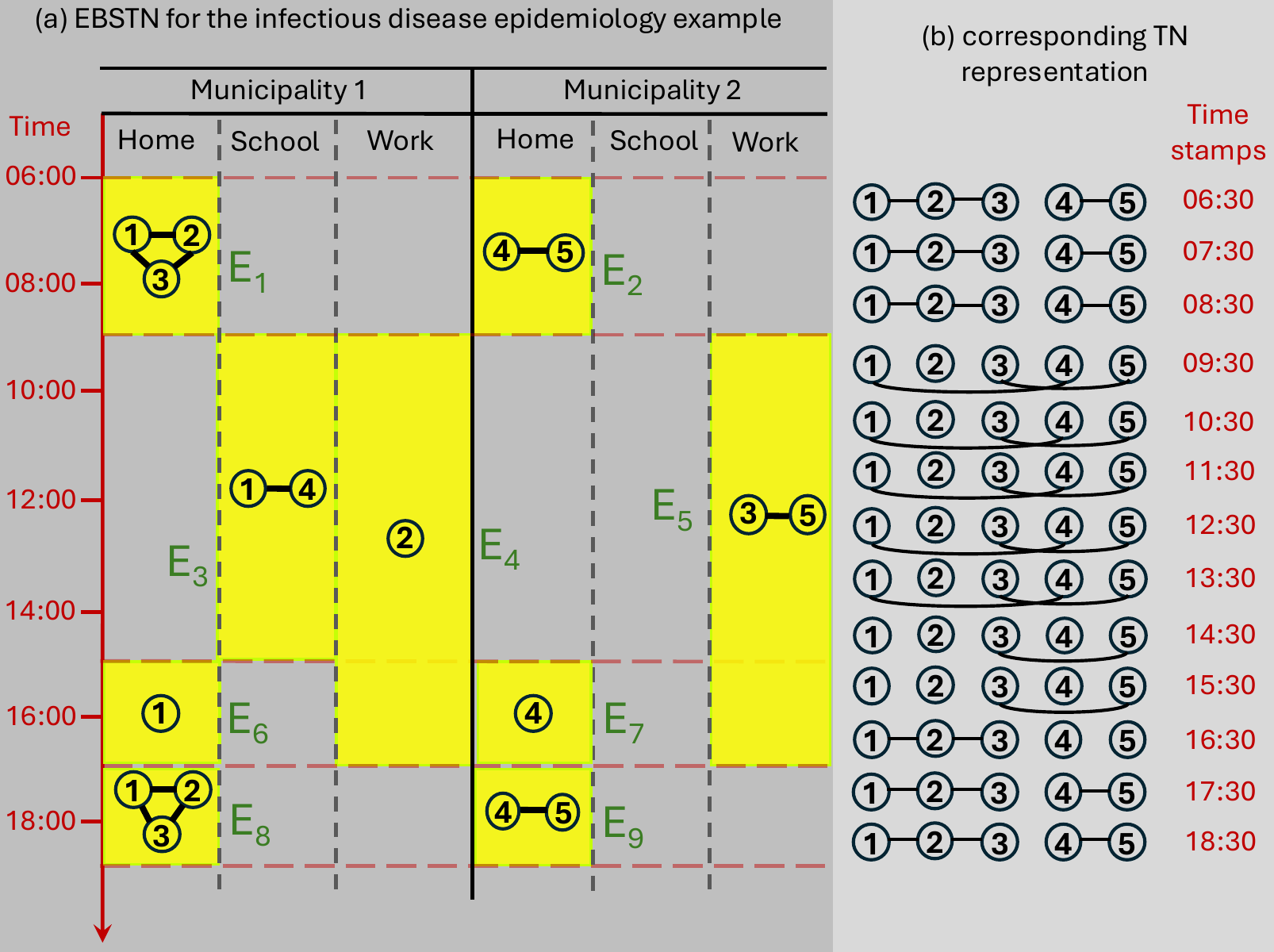}
\caption{{\bf Comparison between EBSTN and the current best, TN, for the infectious disease epidemiology example.} Mixing on a day amongst the family members of two different families in the model (a total of five actors), The families reside in two different municipalities. (a) The event-based-spatio-temporal network (EBSTN): events are shown in the same scheme as Fig. \ref{fig1}(e). By construction, the EBSTN leads to the fragmentation of the full network of actors in terms of fully connected components, which is not the case for the corresponding temporal network (TN) --- formed by taking snapshots at 30 minutes past the hour at hourly interval in panel (b) in the same scheme as \ref{fig1}(c) --- resulting in worse scalability for the latter with the number of actors in comparison to EBSTN. See text for details.}
\label{advantage}
\end{figure}

\noindent We use the infectious disease epidemiology application to illustrate the computational advantage offered by EBSTNs, in comparison to its closest relative, TNs. Modelling of this complex system for the Netherlands in its entirety begins with a mechanistic analysis of the system’s micro-scale constitutive processes: the transfer of the pathogen from an infectious to a susceptible actor taking place in contexts such as homes, schools, and workplaces, each physically located within administrative units (i.e., Dutch municipalities). The analysis step motivates a compartmentalisation of space --- the Netherlands as a country --- in units of municipalities, followed by a further subdivision into contexts. As actors in the model --- Dutch residents --- move across these contexts over time, following simulated hourly schedules informed by demography, residency, and mobility data, some of them are concurrently present in the same context within any given hour. This concurrency then naturally defines an event [Fig. \ref{advantage}(a)], the spatiotemporally-anchored container wherein the micro-scale constitutive processes take place (and can therefore be mechanistically modelled within the container). Because the number of municipalities and contexts do not increase with the number of total actors $N$, encoding events in this way can be performed, e.g., by means of a precomputed lookup table.

\vspace{2mm}
\noindent Comparison with the corresponding TN [Fig. \ref{advantage}(b)], which provides the basis for agent-based models (ABMs), shows that EBSTNs automatically yields a fragmentation --- into fully connected components --- of the TN at every time step (e.g., each hourly interval). Since the information on actor interactions is identical across the two frameworks at every time step, the TN could, in principle, also be used to model the micro-scale constitutive processes per event per time step. However, this would require sifting through all actor contacts --- in the worst case, $N(N-1)/2$ pairs of actor links --- to achieve the same network fragmentation {\it per time step\/}. Compared to the EBSTNs, this introduces a factor of $N^2$ slowdown in the construction of the events needed for the process modelling step. (Indeed, straightforward ABMs, which adopt the TN representation, encounter this computational bottleneck and therefore suffer from limits on their scalability.) In addition, EBSTNs offer another advantage compared to TNs: when the same event (i.e., with an identical actor composition) persists across multiple consecutive time steps, the process-modelling activity can be handled in a single integrated time step rather than as repeated iterations of the same event, providing further computational savings. Further details on ABM vs EBSTN comparison for scaling with $N$ as well as advantages regarding computer's memory usage can be found in SI A. A similar constitutive process analysis leading to the EBSTNs, along with the resulting computational gains for the public transport application, is outlined in the Methods section and further detailed in SI B. 

\vspace{2mm}
\noindent In summary, EBSTNs provide a better scalable and more efficient framework than TNs for modelling how micro-scale heterogeneities shape macro-scale behaviour in complex systems across space and time. A similar approach for public transport, our next illustrative example, is outlined in the Methods section.

\subsection*{Mobility-driven heterogeneities in infectious disease epidemiology}

\noindent Real-world emergent dynamics of infectious diseases in space and time --- our first illustrative example to demonstrate how heterogeneities in the process of pathogen transmission from an infectious individual to a susceptible one at the micro-scale events shape the outcome of epidemic dynamics at the macro-scale --- lacks a natural delineation of spatial and temporal scales. Classical epidemiological models choose to ignore this complexity, and employ frameworks that explicitly or implicitly separate temporal from spatial scales. Examples include compartmental models based on ordinary and partial differential equations \cite{ diekmann2013,kermack1927}, static network models \cite{pastor2001, lloyd2001, newman2002, keeling2005}, and metapopulation models built from ordinary differential equations \cite{gosgens2021, glasser2023}. Such models usually stratify populations into distinct disease states (e.g., Susceptible, Infectious, Recovered), and occasionally incorporate additional covariates. For human pathogens, age is a natural covariate due to differences in transmission intensity (typically higher among children and adolescents) and disease severity (often highest among older adults) \cite{mossong2008, thompson2003}; conversely, pathogens infecting sessile hosts, such as plants or farms, typically require spatial covariates, resulting in integro-differential equation formulations \cite{diekmann1978, vandenbosch1990}. Despite these refinements, 
compartmental and metapopulation models rely heavily on assumptions of homogeneity and uniform mixing within each subpopulation, overlooking heterogeneities that drive the dynamics of real-world disease dynamics.
\begin{figure}[!h]
\includegraphics[width=\linewidth]{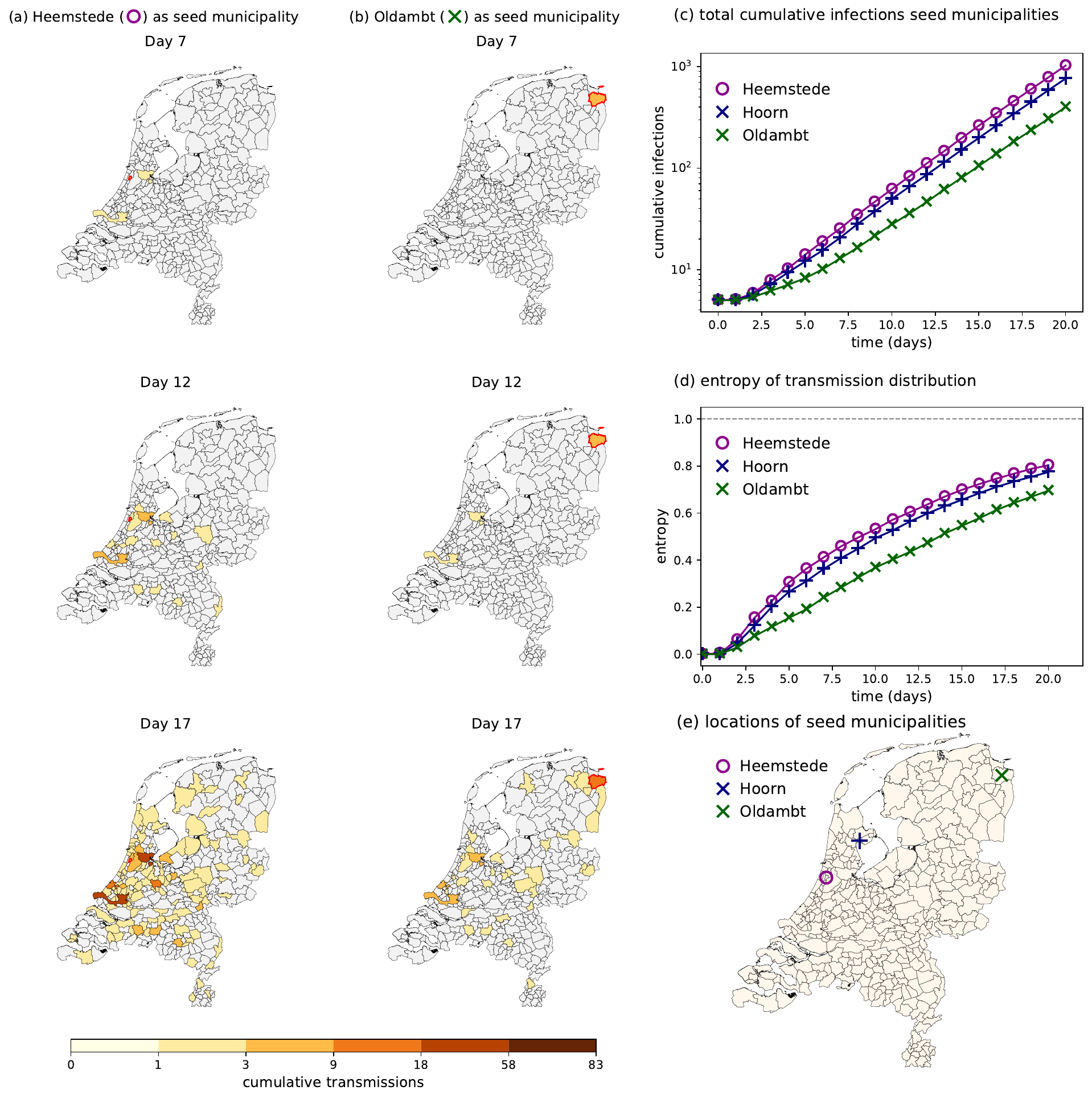}
\caption{Simulations of the early stages of an outbreak of a novel pathogen that is introduced in working adults in selected Dutch municipalities. The model includes $\sim170{,}000$ actors, who are categorised into 11 demographic groups and assigned to $355$ municipalities of residence. Actors' movements are tracked on an hourly resolution, obeying day-night and weekday-weekend mobility rhythms, and they mixed within municipality-specific settings such as workplaces, homes, and schools (see Methods). Panels (a) and (b) show the cumulative number of pathogen transmissions at day 7, 12, and 17 post introduction for a central [Heemstede, panel (a)] and a remote [Oldambt, panel (b)] municipality. The large cities [Amsterdam, Rotterdam, The Hague and Utrecht, the dark-coloured patches for day 17 in panel (a)] contribute disproportionately to the overall infections, both in absolute and relative number of infections. Panel (c) shows that an introduction in a densely populated central municipalities leads to more early infections  than an introduction in a remote municipality. Panel~(d) quantifies spatial heterogeneity by displaying the normalised entropy for the distribution of infectious actors across municipalities over time, where a value of unity indicates a uniform distribution of infectious actors across municipalities (see Methods; initial zero value of normalised entropy corresponds to introduction of the pathogen among the working adult residents of one municipality only). Normalised entropy increases to about 0.8 after three weeks, indicating that spatial heterogeneities in the numbers of infectious actors per municipality have become much smaller by then. \label{fig2}}
\end{figure}

\vspace{2mm}
\noindent Recent developments in epidemiological modelling have largely focused on better integration of the existing models with empirical data. Examples include fitting age-stratified compartmental models to hospital admissions and serological surveys \cite{vanBoven2020, viana2021}, applying kernel-based spatial transmission models to outbreaks of avian influenza, classical swine fever and foot-and-mouth disease \cite{mintiens2003, boender2007, boender2010, boender2014}, and employing network models to study global pathogen spread through airline travel networks \cite{brockmann2013}. In network epidemiology, static network models assume that contact patterns remain fixed over time, which is unrealistic, since human interaction patterns are by nature dynamic. Although dynamic network models are becoming more popular, their application remains limited \cite{masuda2017, liu2022, belikNaturalHumanMobility2011}, as scaling human-to-human contacts to population-level projections requires substantial computation time, as discussed around Fig. \ref{advantage} (see also SI A).

\vspace{2mm}
\noindent In Fig. \ref{advantage}, we have outlined the construction process of EBSTNs, and the computational efficiency they provide in bridging the gap between micro-scale constitutive processes and emergent dynamics at the macro-scale. To illustrate this process in a real-world setting by means of tracking the number of infectious actors at municipality level, we considered the introduction of a respiratory pathogen in a susceptible population in a specified municipality in the Netherlands, thereby extending an earlier work on SARS-CoV-2 \cite{dekker2023} (municipalities are the lowest administrative unit in the Netherlands). Leveraging demography and residency registry data from the Netherlands, we built computer realisations of the hourly movements and interactions of more than $170,000$ actors (corresponding to a $1:100$ population ratio for the Netherlands), adhering to day-night and weekday-weekend travel and interaction patterns for Dutch residents, and accounting for municipality-specific interaction settings such as workplaces, homes, and schools. This generated a large-scale, population-level spatiotemporal contact network among actors, defined at municipality and hourly resolution, on which we simulate the pathogen’s transmission dynamics (see Methods).

\vspace{2mm}
\noindent Fig.~\ref{fig2} shows the cumulative number of pathogen transmissions on this large spatiotemporal network on day 7, 12, and 17 days post introduction for a central [Heemstede, panel~(a)] and a remote [Oldambt, panel~(b)] municipality. The results reveal pronounced spatiotemporal heterogeneities during the early stages of the resulting epidemic. We note that the major urban centres [Amsterdam, Rotterdam, The Hague, Utrecht: dark coloured patches for day 17 in panel (a)] account for a disproportionate contribution to overall transmissions, both in absolute and per capita inhabitants terms \cite{panja2025}. Additionally, an outbreak in densely populated central municipalities leads to a greater overall number of early infections [panel~(c)]. Finally, panel~(d) quantifies spatial heterogeneity by displaying the normalised entropy for the distribution of infectious actors per  municipality over time (see Methods), such that a value of unity indicates a uniform distribution across municipalities, while low values correspond to strong spatial clustering of infections. Normalised entropy gradually increases from zero at the start, reaching approximately 0.8 after about three weeks. This indicates that spatial heterogeneities of infected actors in the early dynamics of the epidemic becomes small on a timescale of approximately three weeks. The dynamics of the normalised entropy in time is impacted by stochastic effects combined with differences in demographic composition of different municipalities \cite{panja2025}. Of note, while the earlier SARS-CoV-2 based analyses used tailored mobility --- i.e., calibrated to observed mobility data --- across municipalities \cite{dekker2023}, here, for the sake of simplicity, we used the gravity model for mobility \cite{simini2021}. The results presented here may therefore represent a lower bound for the degree of spatiotemporal heterogeneity that is expected for a novel pathogen in the Netherlands.

\vspace{3mm}
\noindent Notably, the model is also able to capture the impacts of superspreading on epidemic dynamics at the macro-scale, which we elaborate on in SI C.

\subsection*{Heterogeneous development and spreading of delays in public transport}

\begin{figure*}
\begin{center}
\includegraphics[width=\linewidth]{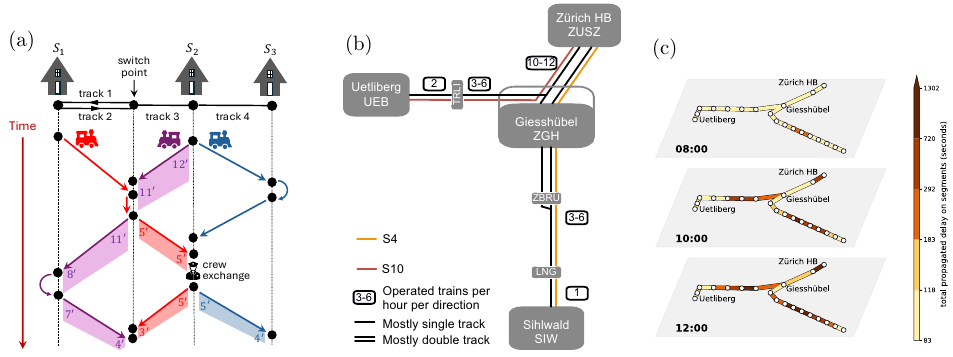}
\end{center}
\caption{{\bf Modelling of public transport system, specifically railway operations.} (a) An illustrative situation with periodic services (see main text) reported as time-distance diagram. A single track is available between the switch point and station $S_2$, and between stations $S_2$ and $S_3$ in Fig. \ref{fig3}(a); once a train enters such a section, no other train can enter it until the occupying train exits. In Fig. \ref{fig3}(a), the scheduled (periodic) services --- and executed ones if there are no delays --- are shown by dark-coloured arrows, while the correspondingly lighter shaded trapezoids indicate the evolution of resulting delays in minutes,  in an operational setting, when an example initial delay is considered. Scheduled services are shown by dark-coloured arrows, each colour identifying a vehicle operating that service. Turnaround of vehicles between two successive services is also reported as a curved arrow of the same colour. An initial delay of 12 minutes occurs for the magenta train departing from  $S_2$. The light-coloured trapezoids indicate linearly-varying delays in minutes along service trajectories. One additional constraints is the crew exchange between the blue and the red service. All tracks are made of multiple `blocking sections' (not displayed for graphical simplicity) arranged as single-track. Because of safety reasons, each blocking can only can be occupied by one train at any time (this is why the red train has to wait for the magenta one to arrive at the switch point before proceeding to track 3). Available buffers allow the delay to reduce throughout the operations. (b) The public transport system in Z\"urich, Switzerland run by operator SZU (Sihltal-Z\"urich-Uetliberg Bahn), comprising two lines with 21 stations in total, on mostly single track infrastructure. (c) The result of the microscopic simulation of delay propagation on the SZU network, using the EBSTN approach. See further elaboration on (c) in the main text.  
\label{fig3}}
\end{figure*}

\vspace{2mm}
\noindent Public transport systems are our second illustrative example, chosen to showcase the precise and scalable handling of discrete dynamics involving multiple moving resources (crew and vehicles). Interactions among actors, e.g., assembled crew and vehicles to run services, are heterogeneous in time, space, intensity, type, duration, and their different time-space trajectories, subjected to constraint events such as shared infrastructure. The emergent phenomena of interest is the spatiotemporal evolution of delays. Public transport systems are time-sensitive, where, constrained by crew and vehicles availability (and in some cases space-constraints), every service  operator strives to adhere to the publicly-available schedule. Due to the strive for efficiency, the resources are heavily utilised, and their binary (yes/no) availability in time and space  strongly constrains the  dynamics of the system. This completely precludes the use of dynamical systems approach for modelling system dynamics, unless all details from microscopic scales are ignored. Instead, graph-theoretical models (Timed Event Graphs or Event Activity Networks) are common in transport literature for modelling of events and resource requirements \cite{pert}, as the two equivalent formulations of  activity-on-arcs or activity-on-nodes \cite{blazewicz2002review}. Because of  their mathematical simplicity, graph-theoretical models are ideal candidates for quantifying the time evolution of (expected) delays and delay-absorbing buffers in a deterministic or stochastic manner \cite{BUKER201234}. When services are periodic, as most timetables prescribe, additional simplifications help identify repeating relationships among activities \cite{GOVERDE2007179}. These models are used in optimisation methods for constructing timetables, e.g., using  `time-distance diagram' [see Fig. \ref{fig3}(a); we chose a railway example for its unique (safety-related) `blocking section' constraints, explained in the Methods section], to determine the resources to be used for an event --- including changing routes, vehicles, crew and event durations --- for improving system performance (e.g., using completion time or delays as cost functions) \cite{MASCIS2002498, SCHOBEL2017348}.   

\vspace{2mm}
\noindent Approaches that do not optimise timetables, instead model the time evolution of a transport system with a given initial condition, can simulate thousands of operations over large networks for multiple hours, if the dynamics is deterministically prescribed. The most advanced of these models use simulated data calibrated to reality, represented by specified probability distributions \cite{BUKER201234}; the distributions could in theory be calibrated differently to individual transport activities, but in practice are assumed to be just a handful of categories depending on the place, type of activity and category of vehicles involved \cite{buchel2020empirical}. When directly using real data, the main roadblock encountered by these approaches stems from the coherent and specific modelling of all types of constraints for the appropriate resources, especially when microscopic details of  blocking sections are required, culminating into bottlenecks pertaining to scalability and incorporation of real-world data.

\vspace{2mm}
\noindent In comparison, EBSTNs offer a unified, simple, and effective way to describe any type of interaction among any number of actors, with varying temporal or spatial characteristics, whether operations follow the timetable or deviate significantly from it. In the full-system EBSTN analogue of Fig. \ref{fig3}(a), each blocking section is treated as a spatial compartment, and each service is an event representing the assembly of the crew and vehicle executing that service in the specified location (see Methods). This structure makes blocking constraints straightforward to implement. EBSTNs are (1) precise, because they model every activity at each blocking section and for each resource (as opposed to static models that ignore exact timing); (2) computationally efficient, because events are represented only by their start and end times (unlike temporally layered networks, such as TN, which replicate the network at fixed time resolutions and link activities across layers); and (3) scalable, because they can be preprocessed to remove redundant constraints and accelerate computation (unlike event activity networks, which must encode all possible sequencing constraints for all infrastructure, vehicle, or crew resources). As a result, EBSTNs support extensions that, while theoretically possible, have not been feasible in current state-of-the-art transport models, enabling the identification of global emergent phenomena through microscopic representations of real-world operations on large transport networks. A comparison of computational efficiency of the state-of-the-art transport models with that of EBSTNs is discussed in SI B.

\vspace{2mm}
\noindent These capabilities are illustrated in Figs. \ref{fig3}(b–c). Figure \ref{fig3}(b) shows a schematic of the SZU railway network in Zürich, Switzerland. Figure \ref{fig3}(c) presents the mean of 10 EBSTN-based simulation runs of delay propagation on this network (see Methods, variations around the mean is shown in SI D). Each of the three temporal layers reports the total propagated delay, measured every two hours, on each segment. Exogenous delays, introduced as noise, were injected into the schedule starting at 06:00, and the EBSTN framework simulated their evolution and propagation, reaching up to ten minutes by 12:00. The three snapshots show how delays, initially absent at 06:00 (not shown), grow and spread across the network. Spatial heterogeneity in delays arises naturally from the blocking-section constraints (see Methods).

\subsection*{EBSTNs in other potential fields of applications}

We now discuss the potential for EBSTNs to address challenges in two different fields of applications: development biology and community ecology.

\vspace{2mm}
\noindent \emph{Development biology}. Traditional approaches to understanding how organs and limbs begin to acquire more complex shapes from an initially flat monolayer sheet of cells have relied primarily on one of two core concepts. Either (1)  contractions on only one side of the tissue sheet create a local preferred curvature~\cite{Pearl2017}, or, (2) interactions with other nearby tissues and/or boundaries provide external forces, in turn either driving shape directly~\cite{Harmansa:2023uq} or setting up mechanical instabilities that can then drive shape~\cite{buckling-nelson, Vellutini2023.03.30.534554}. However, recently it is becoming appreciated that complex patterns of cell behaviour in the plane of the tissue sheet also conspire to produce powerful, tissue autonomous shaping effects~\cite{Fuhrmann2024}. Here, quasi-2D tissue mechanics, dominated by activity along the junctional network connecting cells to their neighbours, give rise to the spontaneous strains required to establish non-trivial shapes in a manner reminiscent of shape-programmable exotic materials, e.g. liquid crystal elastomers or NIPA hydrogels~\cite{ModesWarner2016}. Non coarse-grained, cell-scale models of such tissue mechanics, such as Voronoi Models or Apical Vertex Models~\cite{BiPhysRevX, FLETCHER20142291} minimise an energy functional based on a polygonal mesh network (associated to the cell boundaries), but this alone captures only simple relaxation dynamics. If, on the other hand, topological transitions in the cellular network occur (e.g., neighbour exchanges), these can actively create spontaneous strains and thus drive shape. Unfortunately, such topological transitions do not play well with classical, continuous modelling approaches and currently must either be coarse-grained away or treated in an \textit{ad hoc} manner. Meanwhile, there has been an explosion in high fidelity, high resolution, high field-of-view data of single or collective cell behaviours in epithelial tissues undergoing morphogenesis that cannot be fully addressed by the extant tools. EBSTNs offer an intriguing and powerful option, wherein both the smooth mechanics and topological dynamics can naturally be encoded in the EBSTN and shaping events can be studied and understood in their full complexity. Even better, the chemical signalling environment and internal signalling and regulatory dynamics of the cells could also be captured by EBSTNs, potentially allowing for the simultaneous integration of shaping mechanics together with the dynamical chemical self-organisation that ultimately ``selects'' the shape.

\vspace{2mm}
\noindent \emph{Community ecology}. Events are a central concept in community ecology because they mark interactions --- feeding, competition, pollination, etc. --- between individuals and species. Following the increase in environmental change taking place across the world, it has become clear that the resilience of ecosystems to regime shifts is influenced by multiple and overlapping processes that are highly spatially and temporally dependent~\cite{Scheffer2001}. For example, the shift in dominance from predatory fish to an opportunistic mesopredator in the Baltic Sea is driven by when and where trophic interactions occur in heterogeneously connected and functionally differentiated habitat patches~\cite{Olin2024}. As dispersal distances, rates, and timings vary between the two consumer species, and even between individuals belonging to the same species~\cite{Berkstrom2021}, an accurate picture of network dynamics requires careful tracking of movements among habitat patches. This heterogeneity in both ecological processes and organismal attributes is characteristic of multispecies ecological systems and lends itself well to representation as EBSTNs.

\vspace{2mm}
\noindent In addition to providing a more accurate picture of ecological dynamics, moving to a spatiotemporal network perspective can also inspire more efficient data collection. In community ecology, the bottleneck is often too little data rather than too much data. Rather than sampling at regular time intervals for fixed time periods, switching to collecting data based on when and where important events are likely to occur, mirroring how data are eventually represented in spatiotemporal networks, will optimise collection resources. Even when data are readily available, aligning collection methods and on-the-fly sampling with the model input format will lead to further data-processing efficiencies.

\section*{Discussion}

\noindent Summarising, we have introduced event-based spatiotemporal networks (EBSTNs) as a modelling framework for unravelling how micro-scale heterogeneities shape macro-scale behaviour of complex systems in space and time, especially when dealing with large datasets with highly resolved spatial, temporal and actor information on the constitutive processes. Leveraging real-world data, we demonstrated the effectiveness of EBSTNs with two example complex systems: mobility-driven heterogeneities in infectious disease epidemiology and heterogeneous development and spreading of delays in public transport. For the former, following a local outbreak of a novel respiratory pathogen in the Netherlands, we demonstrated how EBSTNs enabled fine-grained tracking of transmission routes and infection patterns through space, time, superspreaders and superspreading (Fig. \ref{fig2} and SI C). For the latter, we used EBSTNs to analyse the development of secondary delays in the Z\"urich S-bahn system. Further, we discussed the potential for EBSTNs to address challenges in two other fields of applications: developmental biology and community ecology. 

\vspace{2mm}
\noindent We note that discrete event-driven simulation, network-based representations, and process-specific local rules are well-established paradigms in their respective domains. A key innovation lies in their integration into a unified event-based networks framework. Although this integration may appear straightforward, we argue that it does represent a conceptual shift rooted in a {\it dual\/} representation of temporal networks. Standard TN representations – serving as the basis of agent-based models (ABMs) are fundamentally actor-centric: actors are the primary entities, and temporal evolution is constructed by tracking their states and interactions over time. In contrast, our framework adopts an explicitly event-centric viewpoint. Here, spatiotemporally anchored events are linked by actor trajectories across the events.

\vspace{2mm}
\noindent These two perspectives are dual to one another in the sense that interactions in actor-centric models correspond to event linkages in our formulation, while actor trajectories emerge as paths through the event space. {\it Crucially\/} --- and conceptually --- this duality makes the (causal) interdependencies among system processes, their timing, and their sequence explicit rather than implicit. This is key for directly connecting micro-scale interactions to macro-scale dynamics. Beyond this conceptual implication, the event-centric framework is flexible and can be adapted to system-specific constraints, as we have demonstrated for the two worked-out examples. 

\vspace{2mm}
\noindent For the two worked-out examples, we showed how the event structure emerges naturally from their constitutive process analyses. We then built the EBSTNs from these events and simulated the system dynamics on the resulting event networks. In other words, the event networks are constructed before the system processes are simulated. This is not always possible, especially in complex adaptive systems where the future event structure depends on the system’s current state. A simple epidemiological example is when the infected actor 1 in Fig. \ref{advantage} feels unwell at school and decides to go home earlier than 15:00. In such cases, the future event structure cannot be known {\it a priori\/}, requiring the use of conventional ABMs. (A short overview of the current state of ABMs in epidemiology is provided in SI E.) ABMs update the full internal state of the system --- every individual actor --- at each simulation time step, so the simulation advances step by step, unlike EBSTNs, which progress directly from event to event. Even in these situations, process analyses that support spatial compartmentalisation can help enable network fragmentation, as illustrated in Fig.~\ref{advantage}. 

\vspace{2mm}
\noindent We also emphasise that EBSTNs do not require predefined or uniform spatial or temporal resolutions. Instead, these resolutions emerge naturally from the constitutive process analyses. In infectious disease epidemiology, for example, we used contexts --- homes, schools, and workplaces --- within municipalities, the smallest administrative units in the Netherlands, as the natural spatial units. In public transport delay modelling, blocking sections served as the spatial units because of their safety relevance. In developmental biology, individual cells would play this role; in community ecology, habitat patches. These units are inherently heterogeneous in real space.

\vspace{2mm}
\noindent Once spatial units are defined, actor mobility broadly determines the minimal required level of temporal resolution. In the Netherlands, the close proximity of municipalities supports an hourly resolution for an epidemiological model. For the public transport example, the temporal resolution dropped to seconds, reflecting the lengths and the corresponding traverse times of blocking sections.

\section*{Methods}

\subsection*{Infectious disease epidemiology model}

The following steps summarise the core mechanisms underlying our model. These steps yield a full dynamic network of actor contacts, denoted as events. As pointed out in the introduction, An earlier edition of this model was used --- including calibration and validation --- to analyse the first COVID-19 pandemic wave dynamics in the Netherlands \cite{dekker2023}.

\subsubsection*{Step 1: Demographic and residential stratification of the actors}

The model replicates the demographic composition of the Dutch population by using a 1:100 scale representation of the 2019 population (17.34 million), resulting in approximately 170,000 actors. Since each municipality must contain whole-numbered actors in every demographic group, the total number of actors is slightly lower than the scaled population. Using age and occupation data from the Dutch Central Bureau of Statistics (CBS) \cite{cbs_statline_2024}, actors were assigned to one of eleven demographic groups, ensuring that each municipality reflects the observed local distribution of age and occupation as accurately as possible. A breakdown of the demographic groups is shown in SI F.

\subsubsection*{Steps 2 and 3: Actor Mobility}

The model captures real-life travel behaviour by assigning each actor a weekly schedule with recurring daily patterns. Hourly movements were sampled from Dirichlet distributions based on gravity model estimates, with the assumption that most time away from home occurs during mornings and afternoons. Travel time was allocated accordingly, with home time split between the start and end of the day and remaining hours filled with visits to other municipalities in random order.

\vspace{2mm}
\noindent Travel behaviour varies by demographic group. Working adults and students spend about 25\% of their time outside their home municipality, while others spend around 5\%. These differences are incorporated through weighted sampling. Two types of travel are distinguished: frequent trips (e.g., commuting), modelled with a standard gravity formulation, and incidental trips (e.g., leisure), which place less emphasis on distance. Commuting patterns dominate on workdays for students and workers, while all groups rely on incidental travel patterns during weekends or in the absence of regular obligations. For frequent trips to work or school a typical gravity model was used with the gravitational constant $G = \nicefrac12$ to account for double counting for return journeys, i.e., the weight for an actor residing in municipality $m$ to travel to municipality $m'$ on a workday given by
\begin{equation}\label{eq:gravweightsfreq}
    w_{m,m'}^{\mathrm{freq}} = \frac{P(m) P(m')}{2R},
\end{equation}
where $P(m)$ is the population of municipality $m$, and $R$ is the distance between municipalities.
For incidental trips to visit family or for recreation, weights were used with $G = \nicefrac17$, and the square root of distance was used:
\begin{equation}\label{eq:gravweightsinc}
    w_{m,m'}^{\mathrm{inc}} = \frac{P(m) P(m')}{7 \sqrt{R}}.
\end{equation}

\vspace{2mm}
\noindent The resulting mobility patterns align well with empirical data. For instance, the model reflects known proportions of individuals living outside their work or study municipality and produces realistic estimates of time spent away from home (16.1\% versus an observed 16.6\%). Validation against national surveys confirms the plausibility of these movement patterns \cite{Kences2020,CBS2017}. Although the model uses simplified gravity-based mobility instead of more complex data-driven methods, prior comparisons has shown little loss in accuracy, suggesting that key heterogeneity is preserved through demographic detail and structured scheduling.

\vspace{2mm}
\noindent To ensure robustness of the results, five independent mobility networks were generated, and simulations were run across all of them. This ensemble approach reduces sensitivity to specific network realisations and strengthens the reliability of the results.

\subsubsection*{Step 4: Social mixing of the actors}

Heterogeneous contact patterns were modelled using mixing matrices derived from empirical contact data \cite{premetal2017}. These matrices guided hourly contact sampling for each actor, based on their demographic group and current location, following the POLYMOD methodology \cite{mossong2008}. For instance, a working adult outside the home during the afternoon would use the ``work'' mixing matrix, making contacts with primary school children unlikely.

\vspace{2mm}
\noindent In total, four $(11 \times 11)$ matrices were used --- home, work, school, and other --- where each element $C_{ij}$ represents the expected number of contacts between demographic groups $i$ and $j$. The choice of matrix depended on both the setting and time of day, with distinct patterns applied for daytime (08:00–18:00) and nighttime periods. The specific mixing matrices applied are shown in SI F.

\subsubsection*{Step 5: EBSTN formation and pathogen transmission}

Using the actor locations from the schedule, contacts were sampled on an hourly basis using context-specific mixing matrices derived from empirical data \cite{premetal2017}. Separate mixing matrices were used for the contexts: `home', `work', `school', or `other'. Contacts were sampled using the method presented in the POLYMOD study \cite{mossong2008}. In this way EBSTNs were formed for the entire duration of the study, on which we simulated pathogen transmission. The EBSTNs were stochastically built: five different EBSTNs were generated for the results shown in Fig. \ref{fig2}. Per introduction municipality, 15 iterations of the transmission model were run on every EBSTN. Results are therefore averaged over 75 runs per introduction municipality.

\vspace{2mm}
\noindent Pathogen transmission was simulated using a Susceptible-Exposed-Infectious-Recovered (SEIR) framework. Pathogen transmission took place based on within-municipality infection pressure, calculated as the sum of contacts with infectious individuals multiplied by probability of infection per contact \cite{dekker2023}. The infection rate or (also called force of infection) $\lambda$ on an actor was calculated as the number of contacts with infectious actors multiplied by $\beta$, the probability of infection per contact. Thus, infectious actors exerted an infection pressure on each susceptible actor weighted by their mixing. The force of infection $\lambda$ on a susceptible actor from demographic group $g$, at time $t$, in municipality $m$, was calculated using the following equation:  
\begin{equation}\label{eq:infpressure}
    \lambda(g,m,t) = \beta \cdot s(t) \cdot \sum_{g'} n_{g,g'} \cdot \frac{I(g',m,t)}{N(g',m,t)},
\end{equation}
where $g'$ and $s(t)$ represent the demographic groups and the actor's daily activity cycle, $n_{g,g'}$ is the expected number of contacts for an actor in group $g$ with actors from group $g'$, and $\frac{I(g',m,t)}{N(g',m,t)}$ represents the fraction of infectious actors in demographic group $g'$ present in municipality $m$ at time $t$ \cite{dekker2023}. The actor was considered exposed if a random number chosen from a uniform distribution with a unit interval was less than $\lambda(g,m,t)$. Latent and infectious periods were drawn from gamma distributions, with parameters loosely based on influenza A and SARS-CoV-2. For details on the transmission parameters, see Tab. \ref{tab:epi}.
\begin{table}[hpt!]
\centering
\renewcommand{\arraystretch}{1.4} 
\setlength{\tabcolsep}{4pt}

\begin{tabular}{|p{7.5cm}||Q{7.5cm}|}
\hline
\rowcolor{white}
\textbf{Parameter} & \textbf{Value} \\ 
\hline
Probability of infection per contact & 0.135 \\
\hline
Latent period & 2 days ($90\%$ range: 0.7-3.9 days)\\
\hline
Infectious period & 5 days ($90\%$ range: 1.4-10 days)\\
\hline
\end{tabular}
\caption{Transmission parameters for the respiratory pathogen.}
\label{tab:epi}
\end{table}

\subsubsection*{Calculation of normalised entropy in Fig. \ref{fig2}(d)}

The normalised entropy $s(t)$ was calculated using the equation
\[
s(t)=-\frac{1}{\ln M}\sum_{i=1}^M p_i(t)\ln p_i(t),
\]
where $p_i(t)$ is the probability that a randomly chosen infectious actor is a resident of municipality $i$ at time $t$, and $M=355$ is the total number of municipalities in the Netherlands, as specified by Statistics Netherlands (CBS) in 2019 \cite{cbs_statline_2024}. The probability $p_i(t)$ is then the ratio of the number of infectious actors that are resident of municipality $i$ at time $t$ and the total number of infectious actors in the Netherlands at time $t$.

\subsection*{Propagation of delays in public transport}

We simulated delay propagation on the Sihltal-Zürich-Uetlibergbahn (SZU) infrastructure --- a commuter railway network integrated into the Zurich S-Bahn system with its own separate infrastructure. Although the railway network is small, it faces complex operation challenges because it is largely a single-track network, and because it is densely operated. 
\begin{figure}[!h]
    \centering
    \includegraphics[width=0.95\linewidth]{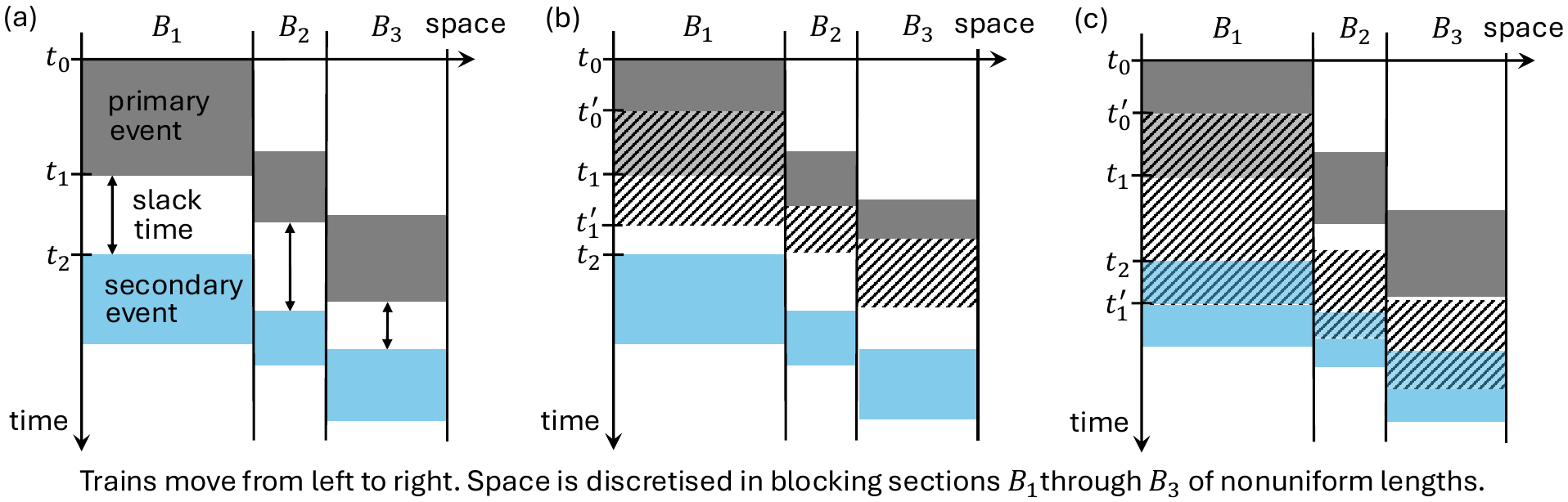}
    \caption{Space-time diagram showing delay propagation on infrastructure (to avoid cluttering the figure, event start and end times are marked only for blocking section $B_1$). (a) The schedule for two different services – preceding (gray) and following (blue) – travelling through three sequential blocking sections of different lengths $B_1\rightarrow B_2\rightarrow B_3$ on an infrastructure link. The event that the preceding (resp. following) service keeps a blocking section occupied we call the primary (resp. secondary) event. Each event --- i.e., running of a service in a blocking section --- requires train unit(s) and crew, i.e., an assembly of these actors of the system (see Fig. SI.1). The duration of the primary event in blocking section $B_1$ is therefore $(t_1-t_0)$. The timespan between the end of the primary event and the start of the secondary event, denoted by double-edged arrows we call the slack time (see text). (b-c) In an actual operation the primary event can be delayed --- shown in hashed rectangles. Events may be delayed at the start, e.g., $(t'_0-t_0)$, and/or pick up further delay during execution, e.g., $(t'_1-t'_0-t_1+t_0)$. (b) The operational end time for the primary event $t'_1$ is smaller than the scheduled start time for the secondary event $t_2$, meaning that no delays propagate to the secondary event. (c) The operational end time for the primary event $t'_1$ is larger than the scheduled start time for the secondary event $t_2$, i.e., $(t'_1-t_2)$ amount of delay propagates to the following service. Evidently, the slack time works as a buffer against delays propagation from the primary service to the secondary service, which is why it is often referred to as the buffer time in literature. The smaller the slack time, the denser the operation.}
    \label{fig:methodstrainsfig}
\end{figure}

\vspace{2mm}
\noindent The simulations used EBSTN framework, for which we first needed to construct the events. As summarised in the main text, construction of the events followed from the analyses of constitutive process --- propagation of delays on railway infrastructure networks, as follows. For safety-related reasons, railway tracks consist of sequentially-placed ``blocking sections'' (SZU has 435 blocking sections): a blocking section at any point of time can only be occupied maximally by one service. How two scheduled services --- a preceding and a following one --- can run on the same track and obey this constraint gets us to the ``space-time diagram'' illustrated in Fig. \ref{fig:methodstrainsfig}(a) \cite{hansen2014}, showing the occupation of the blocking sections by the preceding (resp. following) service for a period of time as the primary event as grey rectangle (resp. secondary event as blue one). The constraint means that the two rectangles cannot overlap within the same blocking section: i.e., there needs to be a ``slack time'' $\Delta t_{\text{slack}}=t_2-t_1$ between them.) 

\vspace{2mm}
\noindent How operational delays of the preceding service can propagate to the following one is shown in Fig. \ref{fig:methodstrainsfig}(b-c). 

\vspace{2mm}
\noindent In a real operational situation, the amount of delay propagated to the secondary service, which is following the primary service, is determined by the primary service's accumulated delay $d_{\mathrm{p,} \mathrm{acc}} = t_0' - t_0$ before the event starts in blocking section $B_1$, and the delay in executing the primary event $\zeta_{\mathrm{p,} \mathrm{runtime}} = t_1' - t'_0-(t_1-t_0)$. The propagated delay $d_{\mathrm{s,} \mathrm{prop}}$ to the secondary service, is then given by
\begin{equation}   d_{\mathrm{s,} \mathrm{prop}} 
= [ d_{\mathrm{p,} \mathrm{acc}}  + \zeta_{\mathrm{p,}\mathrm{runtime}} - \Delta t_{\mathrm{slack}} ]^+.
\label{eq:delayprop1}
\end{equation}
Equation (\ref{eq:delayprop1}) uses Max-Plus notation where $[x]^+ =x$ if $x\ge0$ and $=0$ otherwise. The `p' and the `s' subscripts stand for primary and secondary respectively. This equation can be seen in action in \ref{fig:methodstrainsfig}(b-c). In Fig. \ref{fig:methodstrainsfig}(b) and (c), Eq. (\ref{eq:delayprop1}) translates to $d_{\mathrm{s,} \mathrm{prop}}=0$ and $d_{\mathrm{s,} \mathrm{prop}}=t'_1-t_2$ respectively.

\vspace{2mm}
\noindent To compute delays for further downstream services caused by the secondary service within the same blocking section, one must consider the secondary service’s accumulated delay, which results from the combination of any propagated delay from the primary service and any delay already accumulated by the secondary service at previous blocking sections.

\vspace{2mm}
\noindent Because of such interdependencies amongst events as the constitutive process for delay propagation on infrastructure, an EBSTN approach thus provides the natural path forward for computational modelling of the emergent phenomena (i.e., dynamics of delays in the full SZU system). To this end, a full day of operation was simulated using Eq. (\ref{eq:delayprop1}) for all events in the network. 

\vspace{2mm}
\noindent At each event, $\zeta_{\mathrm{p,} \mathrm{runtime}}$ was chosen, as an exogenous noise term, from the probability distribution $P(\eta)=\tau^{-1}\exp(-\eta/\tau)$, with $\tau = 10$ seconds.

\vspace{2mm}
\noindent As the simulation progressed through the day, the exogenous delay started to accumulate delay over the entire network. During simulation, we measured the propagated delay at each event and attribute it to the location of the secondary event. The blocking section information was then aggregated into reduced infrastructure elements separated by stations, and the corresponding propagated delay averaged over 10 simulations is shown using a stacked heat map in Fig. \ref{fig3} at timestamps 08:00, 10:00, 12:00 hours. The statistics of the delays are summarised in Tab. \ref{tab:transport}.
\begin{table}[hpt!]
\centering
\renewcommand{\arraystretch}{1.4} 
\setlength{\tabcolsep}{4pt}
\begin{tabular}{|p{7.5cm}||Q{7.5cm}|}
\hline
\rowcolor{white}
\textbf{Quantity} & \textbf{Value} \\ 
\hline
\hline
Average primary delay per event & 10 seconds (sampled from an exponential distribution with scale $\tau = 10$s)\\
\hline
Average scheduled slack time & 543.9 seconds, ($90\%$ range: 56.0-1359.0 seconds)\\
\hline
Average duration of scheduled runtime events & 13.9 seconds, ($90\%$ range: 2.0-38.9 seconds)\\
\hline
\end{tabular}
\caption{Details of the transport model, built using the SZU railway network data for one day.}
\label{tab:transport}
\end{table}

\section*{Data Availability}

\noindent The epidemiology and delay propagation simulation data presented in this study have been deposited in the archived Github repository on Zenodo under accession code \url{https://doi.org/10.5281/zenodo.20177183} \cite{NatCommsFigs}. New simulation data can be generated using archived Github repositories: \url{https://doi.org/10.5281/zenodo.20157718} \cite{BlockingSectionsNatCommsPublic}, and 
\url{https://doi.org/10.5281/zenodo.20177177} \cite{covid-simulation-natcomms}.

\section*{Code Availability}\label{sec:code}

\noindent The code used to generate simulation data presented in this study have been deposited in the archived Github repositories on Zenodo under accession codes \url{https://doi.org/10.5281/zenodo.20157718} \cite{BlockingSectionsNatCommsPublic}, and \url{https://doi.org/10.5281/zenodo.20177177} \cite{covid-simulation-natcomms}.

\section*{Acknowledgements}
\noindent We thank Jan Lordieck for providing the schematic figure \ref{fig3}(b), and many helpful discussions on delay propagation in public transport.


\section*{Author contributions}
\noindent DP conceived the idea for EBSTNs. MR carried out all computations. MR, MvB, FC, CM, PS and DP contributed to sharpening the concepts, edited and reviewed the final manuscript.

\section*{Competing Interests Statements}
\noindent The authors declare no competing interests.

\end{document}